\documentclass[aip,prl,twocolumn,longbibliography]{revtex4}
\usepackage[usenames,dvipsnames]{color}
\usepackage{graphicx}
\usepackage{dcolumn}
\usepackage{bm}
\usepackage{amsmath}
\usepackage{amsfonts}
\usepackage{psfrag}
\usepackage{mathtools}
\usepackage{accents}

\usepackage[mathcal]{euscript}

\begin{document}

\title{Nonlocal Response of Polar Nanostructures}

\author{Christopher R. Gubbin}
\author{Simone De Liberato}
\email[Corresponding author: ]{s.de-liberato@soton.ac.uk}
\affiliation{School of Physics and Astronomy, University of Southampton, Southampton, SO17 1BJ, United Kingdom}

\begin{abstract}
Polar dielectric nanoresonators can support hybrid photon-phonon modes termed surface phonon polaritons with lengthscales below the diffraction limit. In the deep sub-wavelength regime the optical response of these systems was recently shown to diverge from that predicted through a standard dielectric description. Recently we developed an analytical, dielectric approach and applied it to spheres and planar heterostructures, reproducing anomalous features observed in experiment and microscopic calculations. In this Letter we develop tools to describe the nonlocal response of polar nanoresonators of arbitrary symmetry. Their validity is verified by comparison to our previous analytical work, before application to new systems. We show that nonlocal energy transfer into matter-like modes in the dielectric diminish field enhancement in nanoscale dimers and that strong nonlocal frequency shifts are possible in macroscopic systems comprised of nanoscopic layers.
\end{abstract}
\maketitle

Nanophotonics is concerned with concentration and control of light on deep-subwavelength scales. This is possible by exploiting kinetic motion of charged particles, allowing the diffraction limit to be beaten many times over \cite{Khurgin2017}. This is the basis for polar nanophotonics, where photons are hybridised with the optic phonons of a crystal lattice in modes termed surface phonon polaritons \cite{Greffet2002,Hillenbrand2002,Caldwell2015a}. These modes are highly tuneable \cite{Spann2016,Gubbin2016,Passler2018, Ellis2016,Gubbin2017} and have broad applications in nonlinear optics \cite{Gubbin2017b, Razdolski2018} and the fabrication of nanophotonic circuitry \cite{Li2016,Li2018,Chaudhary2019}.\\
A key benefit of localised surface phonon resonances (LSPhRs) is their strong morphological dependance. In geometries containing sharp corners, or small gaps this results in a dramatic increase in local energy density which can be used for sensing applications \cite{Berte2018}. When the confinement length approaches the atomic length scale the finite wavelength of the longitudinal optic (LO) and transverse optic (TO) phonons becomes important. LO modes act to push screening charges induced as the particle boundary into the interior. This leads to a divergence from local theories of dielectric response, which assume screening charges are exactly localised at the scatterer boundary. The nonlocal regime has been studied in plasmonic systems, where excitation of strongly evanescent bulk plasma waves leads to a charge smearing which limits maximal field enhancement and leads to blue shifted modal frequencies \cite{Ciraci2013, Mortensen2014}.\\
The nonlocal regime is difficult to access, requiring fabrication of nanoscale resonators or gaps \cite{Ciraci2012}. It is however expected to be of particular interest for phonon polaritons in the field of crystal hybrids, which are constructed from many alternating nanoscale layers of different polar dielectric materials. In these systems recent studies have shown strong divergence from the local optical response \cite{Ratchford2019}. These effects can be modelled using a first-principles method such as density functional theory, \cite{Jones2015, Rivera2018} however these methods scale badly to realistic devices with large numbers of degrees of freedom. To provide more agile methods we recently developed an analytical continuum model, describing polar nonlocality in terms of macroscopic fields \cite{Gubbin2020}. As in local scattering theory analytical models are tractable in systems with strong symmetry but cannot be easily generalised beyond this.\\
In this Letter we develop new numerical tools to describe the nonlocal response of polar nanosystems with arbitrary geometry. This is achieved through integration of our nonlocal response theory with a commercial finite element solver, the model is distributed for the use of the community \cite{Model}. The approach is validated by comparison to our previous analytical nonlocal scattering spectra for 3C-SiC spheres \cite{Gubbin2020}. The model is then applied to the study of spherical dimers, demonstrating the effect of nonlocality on field confinement. Finally we study the nonlocal response of macroscopic resonators containing nanoscale layers, demonstrating the effect of nonlocality in a crystal hybrid resonator.

Our polar crystal is treated in the continuum limit as an isotropic lattice with a single phonon branch characterised by zone-centre longitudinal (transverse) phonon frequencies $\omega_{\mathrm{L}} \; ( \omega_{\mathrm{T}} )$ in a quadratic dispersion approximation analogous to that used in nonlocal plasmonics \cite{Ciraci2013}. Phonons couple to driving electric field $\mathbf{E}$ as
\begin{equation}
	\left[\omega_{\mathrm{T}}^2  - \omega\left(\omega + i \gamma\right) \right] \mathbf{X} + \nabla \cdot \boldsymbol{\tau} - \frac{\mu}{\rho} \mathbf{E} = 0, \label{eq:IonEOM}
\end{equation}
in which $\mathbf{X}$ is the relative ionic displacement, $\gamma$ is the damping rate and $\rho$ and $\mu$ are effective mass and charge densities. The matrix $\boldsymbol{\tau}$ describes the phonon dispersion, acting as an effective stress tensor \cite{Gubbin2020, Trallero-Giner1994a} given for an isotropic lattice by
\begin{equation}
	\boldsymbol{\tau} = \beta_{\mathrm{T}}^2 \left(\nabla \mathbf{X} + \left(\nabla \mathbf{X}\right)^{\mathrm{T}}\right)  + \left(\beta_{\mathrm{L}}^2 - 2 \beta_{\mathrm{T}}^2\right) \nabla \cdot \mathbf{X} \bar{\mathrm{I}}, \label{eq:Stress}
\end{equation}
where $\beta_{\mathrm{T}} \; (\beta_{\mathrm{L}})$ are phenomenological velocities describing transverse (longitudinal) phonon dispersion. The model is completed by the constitutive relation
\begin{equation}
	\mathbf{P} = \mu \mathbf{X} + \epsilon_0\left(\epsilon_{\infty} - 1\right) \mathbf{E}, \label{eq:cons}
\end{equation} 
where $\epsilon_{\infty}$ is the high-frequency permittivity and the material polarisation is $\mathbf{P}$.\\
In a previous publication we solved Eq.~\ref{eq:IonEOM} and Maxwell's equations analytically in simple systems. In principle they could be solved self-consistently by any numerical method. Here we employ a method which permits easy non-uniform meshing as oscillations induced by Eq.~\ref{eq:IonEOM} occur on the nanometer scale, while the wavelength of mid-infrared photons is typically 4 orders of magnitude larger. Non-uniform meshing is simple using commercial finite element (FEM) solvers. To use Eq.~\ref{eq:IonEOM} in a FEM calculation it must be translated into weak form. The strong statement is that the left-hand side of Eq.~\ref{eq:IonEOM} is zero everywhere. The corresponding weak statement is that it, integrated over the computational domain and multiplied by a family of test functions $\boldsymbol{\Phi} $, is zero. Integrating by parts over the computational domain yields
\begin{align}
	\int \mathrm{d^3 r} \biggr[ & \frac{\mu}{\omega_{\mathrm{T}}^2 - \omega \left(\omega + i \gamma\right)}\nonumber \\
	&\times \left( \beta_{\mathrm{L}}^2 \left(\nabla \cdot \mathbf{X}\right) \left(\nabla \cdot \boldsymbol{\Phi}\right) -  \beta_{\mathrm{T}}^2 \left(\nabla \times \mathbf{X}\right) \cdot \left(\nabla \times \boldsymbol{\Phi} \right)\right) \nonumber \\
	& + \alpha \mathbf{X}\cdot \boldsymbol{\Phi} + \epsilon_0 \left(\epsilon_{\mathrm{LRA}}\left(\omega\right)  - \epsilon_{\infty}\right) \mathbf{E} \cdot \boldsymbol{\Phi} \biggr] = 0,\label{eq:weakEoM}
\end{align}
where we simplified using the local dielectric function
\begin{equation}
	\epsilon_{\mathrm{LRA}} \left(\omega\right) = \epsilon_{\infty} \left[ 1  + \frac{\alpha^2/\epsilon_0 \epsilon_{\infty} \rho}{\omega_{\mathrm{T}}^2 - \omega \left(\omega + i \gamma\right)}\right]. 
\end{equation}\\
In coupling Eq.~\ref{eq:weakEoM} with Maxwell's equations we introduce the macroscopic fields, $\mathbf{X}$ and $\boldsymbol{\tau}$. This means the Maxwell boundary conditions are insufficient to determine mode amplitudes in each layer. The appropriate boundary conditions can be derived considering energy transport across a material interface \cite{Nelson1996, Gubbin2020}, these are continuity of the normal and in-plane ionic displacement $\mathbf{X}$ and of the normal and shear components of the effective stress $\boldsymbol{\tau} \cdot \hat{\mathbf{n}}$. At interfaces between local and nonlocal layers the appropriate combination of boundary conditions has been the subject of active debate \cite{Rimbey1974, Pekar1959, Ridley1992}. It is necessary to apply two of the four conditions, leaving a discontinuity in the remaining components. In plasmonic nonlocality the surface normal current is constrained \cite{Raza2011, Mortensen2014, Ciraci2013}, we follow that convention here by fixing the normal displacement. The second additional condition is addressed later in this Letter.\\
\begin{figure}
	\includegraphics[width=0.5\textwidth]{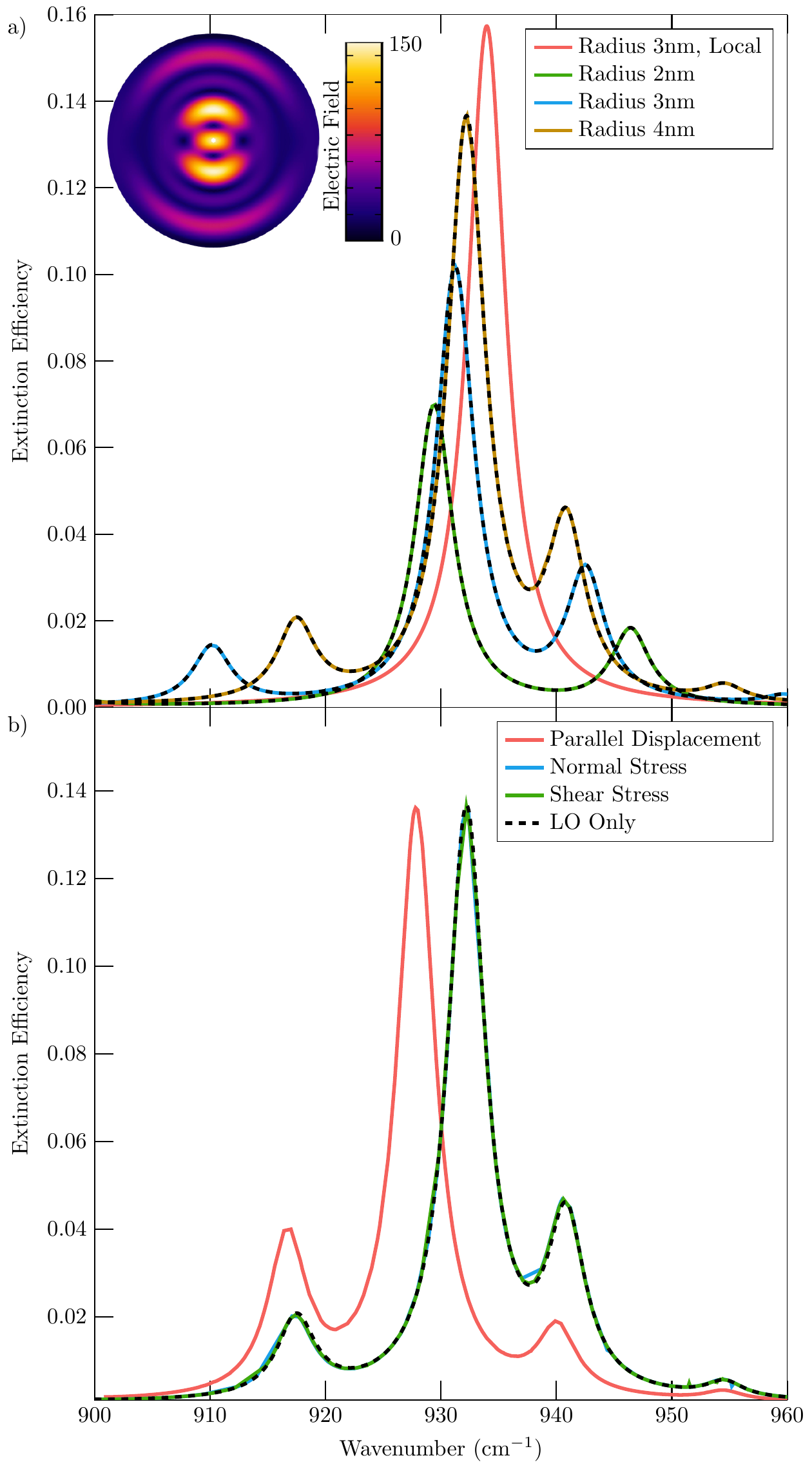}	
	\caption{\label{fig:Fig1} a) Numerical and analytical extinction cross sections for spheres of radius $r=4, 3, 2$nm. The numerical results are shown by solid lines. All analytical results are shown by black dashed lines. The local result is also shown for $r=3$nm. Inset shows the nonlocal electric field magnitude for $r=4$nm. b) Comparison of numerical extinction cross sections using various additional boundary conditions for $r=4$nm.}
\end{figure}
To verify our model we study a system whose nonlocal response is analytically calculable. We demonstrated that nonlocal extinction spectra of nanoscopic 3C-SiC spheres are well described by a quasistatic model in which the transverse dispersion is neglected and only the boundary condition on the normal displacement is enforced \cite{Gubbin2020, Raza2011}. This is a reasonable approximation near the Fr{\"o}hlich resonance where the transverse phonon is strongly evanescent. Considering the case $\beta_{\mathrm{L}} = 15.39 \times 10^5  \mathrm{cm \; s}^{-1}$ \cite{Karch1994} we calculate nonlocal extinction efficiencies.\\
 Results are shown in Fig.~\ref{fig:Fig1}a for radius $r = 2, 3, 4$nm. In the local case small spheres exhibit a single resonance at the Fr{\"o}hlich frequency, $\omega_{\mathrm{F}} \approx 933 \mathrm{cm}^{-1}$, illustrated for $r=3$nm by the red curve. In the nonlocal case additional peaks appear in the extinction spectrum. These  correspond to quantized LO phonon modes. Eventually the Fr{\"o}hlich resonance red shifts as a result of an increase in the effective nonlocal dielectric function. Note that in Fig.~\ref{fig:Fig1}a analytical (numerical) results are illustrated by solid (dashed) lines, overlap is exact on this scale  demonstrating the accuracy of our implementation. The inset shows the nonlocal electric field magnitude for the 4nm sphere, the short wavelength LO phonon oscillation is clearly visible.\\
Eq.~\ref{eq:IonEOM} is a continuum approximation, treating the phonon dispersion phenomenologically through effective stress tensor Eq.~\ref{eq:Stress}. Phonon dispersions are assumed quadratic, meaning LO and TO dispersion relations have solutions at all frequencies. In reality the granular structure of the lattice prevents this, resulting in a decrease in group velocity to zero at the Brillouin zone edge \cite{cohen_louie_2016}. In 3C-SiC TO dispersion is weak, meaning that using finite $\beta_{\mathrm{T}}$ should not alter the extinction cross-section.\\
We study this in Fig.~\ref{fig:Fig1}b for a $r=4$nm and $\beta_{\mathrm{T}} =  9.15 \times 10^5 \; \mathrm{cm \; s}^{-1}$ \cite{Karch1994}, for each of the three secondary boundary conditions. Results are shown in Fig.~\ref{fig:Fig1}b. Results constraining shear or normal components of $\boldsymbol{\tau}$ overlap with the $\beta_{\mathrm{T}} \approx 0 \; \mathrm{cm \; s}^{-1}$ limit explored in the prior section, while that constraining the in-plane displacement is shifted to the red.  For this reason we constrain the normal stress at the scatterer boundary.\\
\begin{figure}
\includegraphics[width=0.5\textwidth]{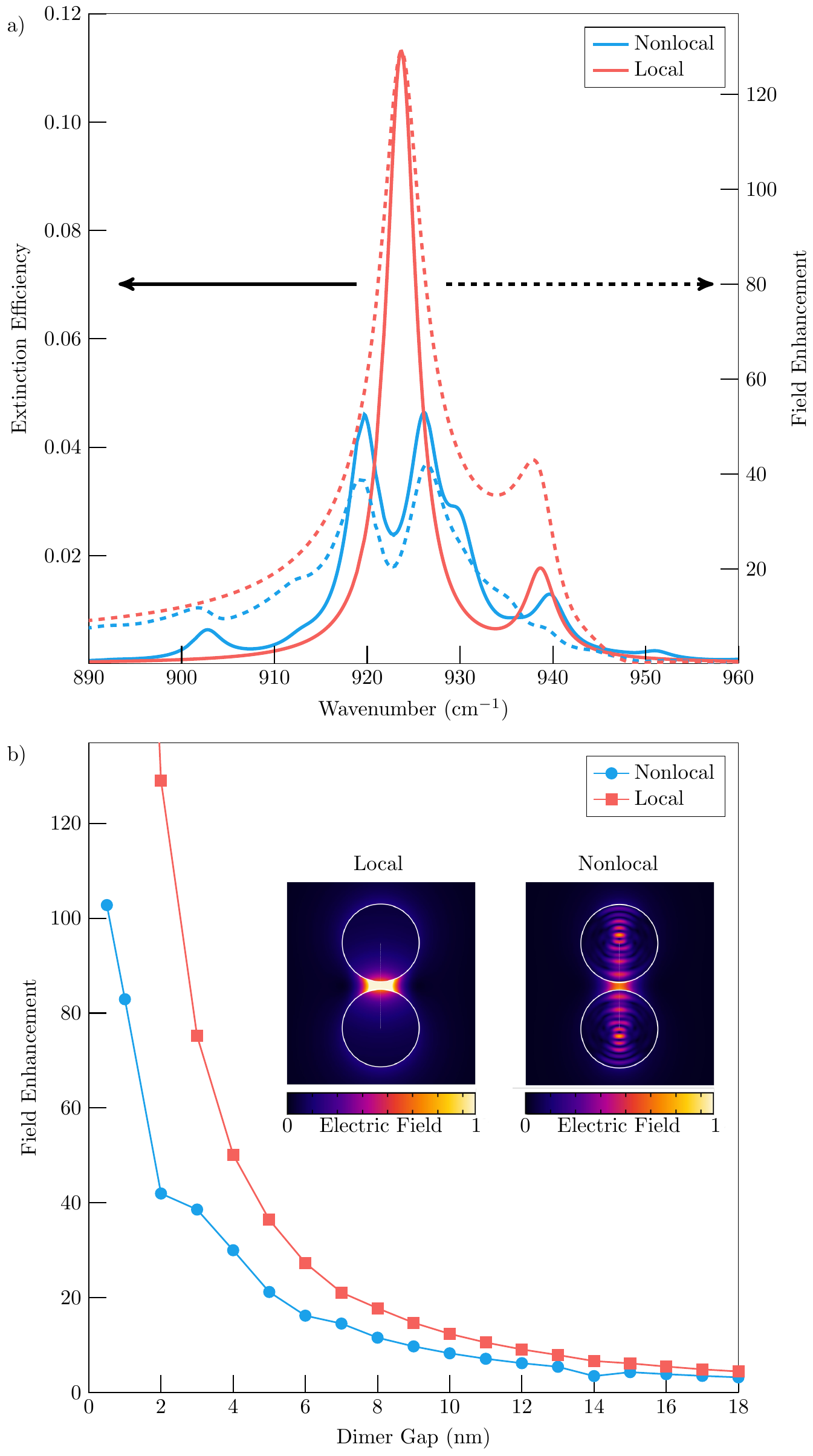}	
	\caption{\label{fig:Fig2} a) Comparison of local (red lines) and nonlocal (blue lines) extinction efficiencies (solid lines) and field enhancements (dashed lines) at gap centre and for a dimer parameterised by sphere radius $r = 5$nm and intersphere gap $d = 2$nm. b) Comparison of nonlocal (red squares) and local (blue spheres) gap centre field enhancements for a dimer of sphere radius $\mathrm{r} = 5$nm as a function of gap width. Lines are provided as a guide to the eye. Inset shows the electric field magnitude of the antibonding resonance for $d=1$nm.}
\end{figure}\\
We have applied our numerical model to systems with analytical solutions, demonstrating its reliability. In the remainder of this Letter we apply it to nanophotonic systems relevant for modern surface phonon polaritons \cite{Caldwell2013, Gubbin2016}, where lack of symmetry prevents analytical solutions. Firstly we study the effect of nonlocality on field hotspots, predicted in plasmonic systems to result in strong charge smearing and a corresponding decrease in the maximal field \cite{Fernandez-Dominguez2012}. We consider a spherical dimer, consisting of two 3C-SiC spheres of radius $r = 5$nm, separated by a gap of width $d$. For large gaps the system modes are those of the isolated spheres studied in Fig.~\ref{fig:Fig1}. For small $d$ these hybridise into bonding and antibonding resonances \cite{Prodan2003}, as shown in the local scattering spectra for $d = 2\mathrm{nm}$ in Fig.~\ref{fig:Fig2}a at $918 \mathrm{cm}^{-1}$ and $935 \mathrm{cm}^{-1}$ respectively. In the nonlocal spectra (solid line) these modes are supplemented by the longitudinal modes supported by the dimer, as in Fig.~\ref{fig:Fig1}a.\\
The antibonding mode is of most interest as opposing charges enclosing the gap result in strong capacitative field enhancement. This is demonstrated in the local case by the dashed line in Fig.~\ref{fig:Fig2}a, which shows field enhancement at gap centre. On resonance in the local case this peaks at around $125$. In the nonlocal case the enhancement diminishes to around $40$. This can be understood from the field intensity plots inset in Fig.~\ref{fig:Fig2}b. In the local case field is strongly localised in the gap, and is efficiently screened from the sphere interior. In the nonlocal case screening is less efficient and induced screening charges smear into the spheres, diminishing capacitative charging of the dimer. Gap-dependance is demonstrated in Fig.~\ref{fig:Fig2}b, where we plot peak field enhancement for the antibonding mode. In the local case field enhancement diverges as $d \to 0$, while in the nonlocal case this is offset by increased energy transfer to propagative LO modes in the nanosphere.\\
\begin{figure}
	\includegraphics[width=0.5\textwidth]{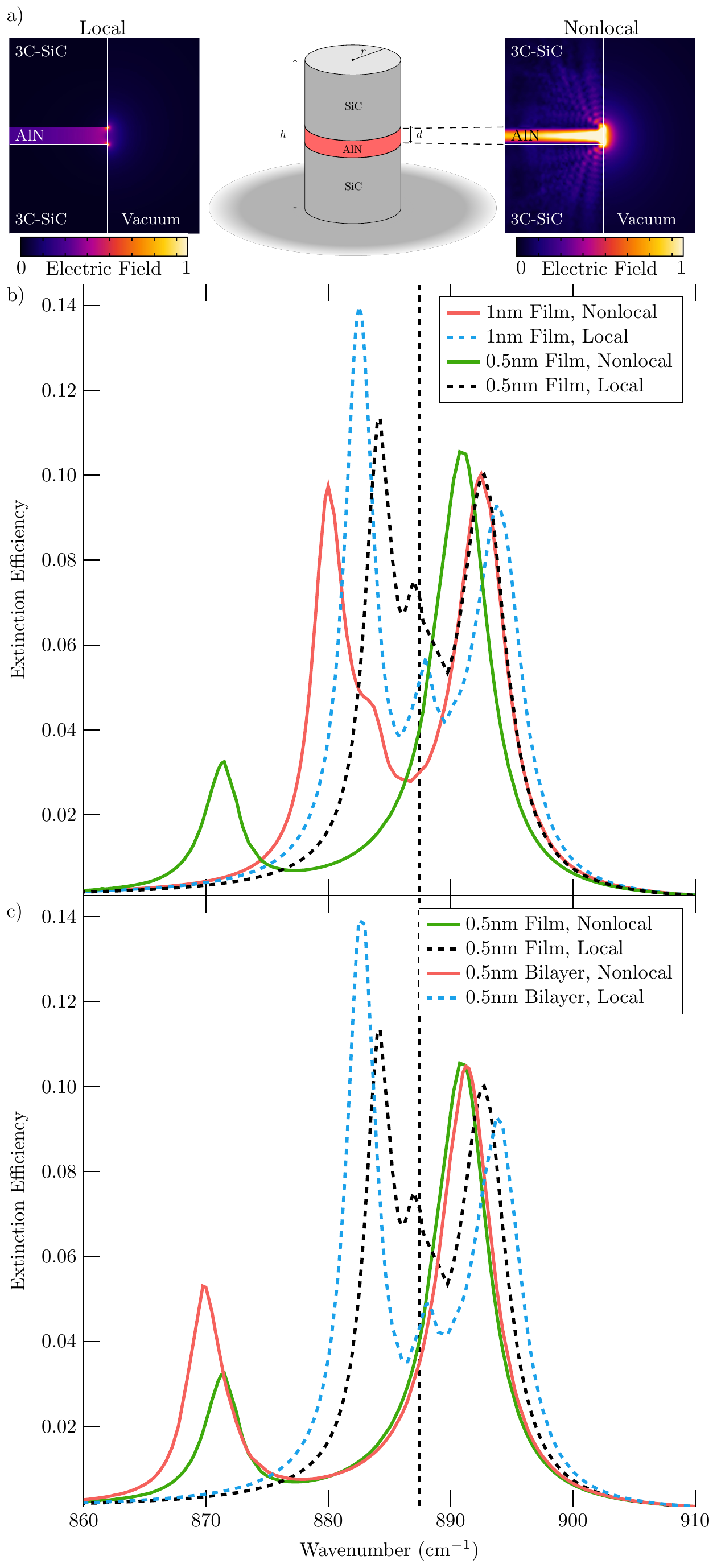}	
	\caption{\label{fig:Fig3} a) Electric field intensity at the nanopillar boundary. b) Comparison of local (dashed lines) and nonlocal (solid line) extinction efficiencies for a 3C-SiC nanopillar of height $250$nm and diameter $500$nm on deep 3C-SiC substrate containing a thin AlN film of thickness $0.5$ and $1$nm. c) Comparison of local (dashed lines) and nonlocal (solid line) extinction efficiencies for a 3C-SiC nanopillar of height $250$nm and diameter $500$nm on deep 3C-SiC substrate containing one and two thin AlN films of thickness $0.5$nm}
\end{figure}
We have discussed nonlocality in systems of nanoscopic dimensions or with nanoscale air gaps. Fabrication of polar resonators on this scale is challenging, however making macroscopic heterostructures containing nanolayers is a well established process. It was recently suggested that in bulk polar superlattices, termed crystal hybrids \cite{Ratchford2019}, a nonlocal description of the optical response is necessary \cite{Gubbin2020}. Describing a system containing many polar layers is beyond the scope of this work, however we demonstrate the effect of nonlocality in larger resonators containing a small number of nanoscopic polar layers.\\
We apply our model to a typical polar resonator, a 3C-SiC nanopillar of height $h=250$nm and radius $r=500$nm on same material substrate \cite{Gubbin2016, Caldwell2013}. We consider a single AlN layer in the pillar centre as illustrated in Fig.~\ref{fig:Fig3}a, parameterised by $\beta_{\mathrm{T}} = 1 \times 10^{5} \mathrm{cm} \; \mathrm{s}^{-1}, \beta_{\mathrm{L}} = 5.1 \times 10^{5} \mathrm{cm} \; \mathrm{s}^{-1}$ \cite{Gubbin2020}. In the dashed lines of Fig.~\ref{fig:Fig3}b we show the local extinction cross section for film thicknesses $d = 0.5, 1 \mathrm{nm}$. The spectrum shows three features. The first, closely resonant with the zone-centre LO phonon in the AlN ($ \omega_{\mathrm{L}} \approx 887 \mathrm{cm}^{-1}$, marked by the vertical dashed line) is the Berreman mode of the AlN film \cite{Passler2018}, independent of the pillar dimensions. Other peaks are photonic modes resulting from hybridisation of the monopolar mode of the nanopillar \cite{Gubbin2016} with the epsilon-near-zero response of the AlN \cite{Passler2018}. In the $1$nm case these are more strongly split around $\omega_{\mathrm{L}}$ as the increased film thickness results in enhanced pillar-film coupling. In the nonlocal case (solid lines in Fig.~\ref{fig:Fig3}b) both photonic modes are red shifted, this is particularly true for the low frequency mode. The red shift is more pronounced for the $0.5$nm film. This is because the AlN film supports quantised Fabry-P{\'e}rot LO phonon modes which red shift as $d \to 0$, resulting in a red shift of the hybridised resonances.\\
Crystal hybrids are comprised not of a single nanoscopic layer embedded in bulk but of many such layers \cite{Ratchford2019}. To demonstrate how these results of can be extrapolated to such systems we consider the effect of adding a second $0.5$nm AlN film, separated from the first by a $0.5$nm 3C-SiC spacer layer, results are shown in Fig.~\ref{fig:Fig3}c. In the local (dashed lines) cases the additional layer results in an increased splitting of the photonic modes around $\omega_{\mathrm{L}}$ as a result of increased coupling between the AlN epsilon-near-zero mode and the pillar resonance. In the nonlocal case the same thing happens around $\omega_1$. A full crystal hybrid would exhibit even larger coupling. Also shown in Fig.~\ref{fig:Fig3}a are the electric field magnitude at the nanopillar boundary in for a 1nm AlN film, normalised to the field maximum in each case. In the local case the field is localised at the pillar edge. In the nonlocal case energy is transferred into the quantised LO mode, with a node at the film centre.\\ 

We presented a numerical method to study the nonlocal response of nanoscopic polar resonators. This extends previous analytical studies, generalising them to study resonators of arbitrary geometry. Furthermore our work on macroscopic nanopillars can be easily extrapolated to describe the optical response of a crystal hybrid nanopillar, containing a large number of nanoscopic layers. The nonlocal response of this system will not be more complex than the results presented in Fig.~\ref{fig:Fig3}. The dominant effect will be an increase in the collective coupling strength between the phonons confined in the AlN and the photonic mode of the pillar, resulting in increased splitting around the zone centre LO mode. This presents a vital first step toward design of crystal hybrid nanoresonators.

\bibliographystyle{naturemag}	
\bibliography{bibliography}
\end{document}